\documentclass[pop,superscriptaddress,amsmath,twocolumn]{revtex4}  %% REVTeX 4.0
\usepackage{graphicx,dcolumn,bm,amsmath}
\usepackage[normalem]{ulem}

\usepackage[dvipsnames,table]{xcolor}
\usepackage{color}
\newcommand{\beq}{\begin{equation}}
\newcommand{\enq}{\end{equation}}

\newcommand\I{\mbox{i}}

\newcommand{\e}{\eqref}

\newcommand \bfr{{\bf r}}

\newcommand{\p}{\partial}

\begin{document}

\title{
Exact solutions for nonlinear development of Kelvin-Helmholtz
instability  for counterflow of superfluid and normal components of Helium II}

\author{Pavel M. Lushnikov}
\email{plushnik@math.unm.edu}
\affiliation{Department on Mathematics and Statistics, University of New Mexico, New Mexico 87131, USA}
\affiliation{Landau Institute for Theoretical Physics, 2 Kosygin St., Moscow 119334, Russia}

\author{Nikolay M. Zubarev}
\email{nick@iep.uran.ru}
\affiliation{Institute for Electrophysics, Ural Branch, Russian Academy of Sciences, Yekaterinburg, 620016 Russia}
\affiliation{Lebedev Physical Institute, Russian Academy of Sciences, Moscow, 119991 Russia}
\date{%Printed
\today
%November 19, 2003
}

\begin{abstract}
A relative motion of the normal and superfluid components of Helium
II results in Kelvin-Helmholtz instability (KHI) at their common
free surface. We found the exact solutions for the nonlinear stage
of the development of that instability. Contrary to the usual KHI of
the interface between two fluids, the dynamics of Helium II free
surface allows decoupling of the governing equations with their
reduction to the Laplace growth equation which has the infinite
number of exact solutions including the formation of sharp cusps at free surface in a finite time.
\end{abstract}

%\pacs{????52.65.-y, 52.65.Ff, 52.38.-r, 52.35.-g}

\maketitle

Kelvin-Helmholtz instability (KHI)  is perhaps the most important hydrodynamic instability which commonly occurs either at the interface
between two  fluids moving with different velocities or for the shear flow of the same fluid \cite{LandauLifshitzHydrodynamics1989}.   Often KHI instability assumes the interface of two
ideal fluids with the tangential jump of the velocity at the interface while finite fluid viscosities suppress the tangential velocity jump.
The nonzero
viscosities of two fluids implies the shear flow near the interface instead of the tangential velocity discontinuity
which somewhat complicates the analysis resulting in qualitatively different Miles
instability \cite{MilesJFMPartI1957,LushnikovAtm1998rus}.

Recently KHI and relative motion of fluids attracted  significant
experimental and theoretical attention in superfluids including
interface between different phases of     $^3$He
\cite{VolovikJETPLett2002,BlaauwgeersEltsovVolovikEtAlPRL2002,VolovikBook2003,FinneEltsovKopninVolovikEtAlRepProgrPhys2006,VolovikUspekhiFizNauk2015} and relative motion of components of
 $^4$He \cite{HanninenBaggaleyPNAS2014,RemizovLevchenkoMezhovDeglinJLowTempPhys2016,BabuinLvovEtAlPRB2016,GaoVinenEtAlJETPLett2016,GaoVinenEtAlJLowTempPhys2017}.  Here we focus on KHI in   $^4$He with free surface in the superfluid phase (He-II state)  which has counterflow of  superfluid
   and normal fluid components \cite{LandauLifshitzHydrodynamics1989} with densities $\rho_s$ and $\rho_n$, respectively.  %The absence of the viscosity in the superfluid
%component  ensures that there is no
%tangential stress at the free surface which simplifies
%the analysis because it allows the jump of the tangential component of velocity %on the free surface.
We assume that both components are incompressible, $\rho_s\equiv const,$ $\rho_n\equiv const $ with the total density $\rho\equiv \rho_s+\rho_n$. The relative motion of these components results in KHI \cite{KorshunovEuriphysLett1991,KorshunovJETPLett2002} which  is somewhat unusual because   that motion  occurs from the same side of the He-II free surface. Besides that, He-II provides   the excellent ground to study the classical KHI (e.g. contrary to winds over oceans) because the tangential velocity jump between two fluid components at the free surface is allowed to satisfy the hydrodynamic equations by the absence of the viscous stress in the superfluid component.

In weakly nonlinear approximation the two-dimensional dynamics of
interface between  two fluids  can be reduced to the motion of
complex singularities through the analytical continuation into
complex plane from the interface. Approach of singularity to the interface means a formation of a cusp. Examples includes dynamics of the
interface of two ideal fluids
~\cite{KuznetsovSpektorZakharovPhysLett1993,KuznetsovLushnikov1995,Zubarev_Kuznetsov_JETP_2014},
the dynamics of the interface between ideal  fluid and light viscous
fluid \cite{LushnikovPhysLettA2004} and bubble pinch-off
\cite{TuritsynLaiZhangPRL2009}. In these  systems the dynamics is
determined by  poles/branch cuts in the complex plane. Extending
such weakly nonlinear solutions into strongly nonlinear  solution is
challenging and mostly was done for the particular case of free
surface hydrodynamics (i.e. the density of the second fluid turns
into zero). Typical singularities in that case are square roots
\cite{TanveerProcRoySoc1991,TanveerProcRoySoc1993,CowleyBakerTanveerJFM1999,DyachenkoLushnikovKorotkevichJETPLett2014}
%DyachenkoLushnikovKorotkevichPartIStudApplMath2016}
which however can be located in an infinite number of  sheets of
Riemann surface through an infinite number of nested square roots
singularities \cite{LushnikovStokesParIIJFM2016}.     Another
exception is the ideal fluid pushed through viscous fluid in a
narrow gap between two parallel plates (Hele-Shaw flow) where an
infinite set of exact solution, some of them involving multiple
logarithms, can be constructed by solving the Laplace growth equation
(LGE)
\cite{Kochina_1945,GalinDoklaAkadNauk1945,ShraimanBensimonPRA1984,HowisonSIAMJApplMath1985,BensimonKadanoffLiangShraimanTangRevModPhys1986,MineevDawsonPRE1994}.

Linear results on KHI are well-known
\cite{LandauLifshitzHydrodynamics1989} while here we focus on
nonlinear regime of KHI and find the exact strongly nonlinear
solutions for the dynamics of the He-II free surface through the
exact reduction to the solutions of LGE. These new solutions, in
particular, describe the formation of cusps (dimples) on He-II free surface in a finite time with both a surface curvature and velocities of components of He-II
diverge at singular points. We   estimate   physical parameters for the
experimental observation of obtained exact solutions.

Superfluid component of He-II allows quantized vortices which are
generated primary from   the container walls \cite{FeynmanBook1955}.
Here we consider the dynamics He-II well away from the walls without
quantum vortices as well as we ignore a vorticity of normal
component  similar to Ref. \cite{KorshunovJETPLett2002} which refers to that approximation as non-dissipative two-fluid description. In
that approximation the dynamics of both fluid components is
potential one,  i.e. ${\bf v}_s=\nabla \Phi_{s}$ and  ${\bf
v}_n=\nabla \Phi_{n}$, where ${\bf v}_s$,  ${\bf v}_n$ are
velocities of superfluid and normal components with  $ \Phi_{s}$ and
$  \Phi_{n}$ being the corresponding velocity potentials.
Incompressibility implies Laplace equation for each component,
$\nabla^2\Phi_{n,s}=0$. We focus on two-dimensional flow $\bfr\equiv
(x,y)$, where $x$ and $y$ are horizontal and  vertical coordinates,
respectively. We assume that both fluids occupy the region
$-\infty<y\le \eta(x,t),$, where $y=\eta(x,t)$ is the
free surface elevation   with the   unperturbed surface  given by
$\eta(x,t)\equiv0$. The flow of both components deep inside He-II
($y\to-\infty$) as well as at $|x|\to \infty$ is assumed to be
uniform following  $x$ direction, which implies $\Phi_{n,s}\to
V_{n,s}\,x$, where $V_{n,s}$ are the corresponding horizontal
velocities. We use the reference frame of the center of mass such
that  $\rho_n V_n+\rho_s V_s=0$ and introduce the relative velocity
$V=V_s-V_n>0$ between fluid components meaning that
$V_{n,s}=\mp\rho_{s,n}V/\rho$.

The dynamic boundary condition (BC) at the free surface  ($y=\eta$) follows from the generalization of Bernoulli Eq. into two fluid components (see e.g. Chap. 140 of Ref. \cite{LandauLifshitzHydrodynamics1989} and Refs.
\cite{KorshunovEuriphysLett1991,KorshunovJETPLett2002})
\begin{align}\label{bern}
\left .
\rho_n\left(\frac{\partial\Phi_n}{\partial t}
+\frac{(\nabla\Phi_n)^2}{2}\right)
+\rho_s\left(\frac{\partial\Phi_s}{\partial t}
+\frac{(\nabla\Phi_s)^2}{2}\right) \right |_{y=\eta}
\nonumber\\
=\Gamma-P_{\alpha}-P_g,
\end{align}
where $P_\alpha=-\alpha\frac{\p }{\p x}[\eta_{x}(1+\eta_x^2)^{-1/2}]$   is the jump of pressure at the free surface compared with the zero pressure outside of fluid,
 $\alpha$ is the surface tension coefficient, $\eta_x\equiv \p\eta /\p x$,  $P_g=\rho g\eta$ is the gravity pressure (the contribution of the acceleration due to gravity
  $g$) and $\Gamma=\rho_n\rho_s V^2/(2\rho)$ is the Bernoulli constant which ensures that Eq. \e{bern} is satisfied at $|x|\to \infty$.

The kinematic BCs at the free surface are given by
\begin{equation}\label{kin}
\eta_t({1+\eta_x^2})^{-1/2}
=\partial_n\Phi_n|_{y=\eta}=\partial_n\Phi_s|_{y=\eta},
\qquad
\end{equation}
where $\eta_t\equiv\p\eta /\p t$,   $\partial_n\equiv {\bf n}\cdot \nabla$ is the outward normal derivative to the free surface with   ${\bf n }=(-\eta_x,1)(1+\eta_x^2)^{-1/2}.$ Eqs. \e{bern} and \e{kin} together with   $\nabla^2\Phi_n=\nabla^2\Phi_s=0$ and BC at infinity form a closed set of equations of two-fluid hydrodynamics for KHI problem.

It is convenient to introduce the average velocity
${\bf v}=(\rho_n{\bf v}_n+\rho_s{\bf v}_s)/\rho$ and the auxiliary potentials\[
\Phi=(\rho_n\Phi_n+\rho_s\Phi_s)/\rho,
\qquad
\phi=\sqrt{\rho_n\rho_s}(\Phi_n-\Phi_s)/\rho
\]
which are linear combinations of $\Phi_n$ and $\Phi_s$ thus both
satisfying  Laplace equation together with $\nabla \Phi=\bf v$. BCs at either $y\to-\infty$ or $|x|\to \infty$ are
reduced to
\begin{equation} \label{bcPhiinfinity}
\Phi\to 0
\end{equation}
and
\begin{equation} \label{phiBCinfinity}
\phi\to-Vx\sqrt{\rho_n\rho_s}/\rho.
\end{equation}
Eq. \eqref{bern} turns into
\begin{equation}\label{bern2}
\left.\frac{\partial\Phi}{\partial t}+\frac{(\nabla\Phi)^2}{2}
+\frac{(\nabla\phi)}{2}^2\right|_{y=\eta}=\frac{c^2}{2}-\frac{P_{\alpha}+P_g}{\rho},
\end{equation}
where $c=\sqrt{2\Gamma/\rho}$ is the constant which has the dimension of velocity. Eqs. \eqref{kin} are reduced to\begin{equation}\label{kin2}
\eta_t(1+\eta_x^2)^{-1/2}
=\partial_n\Phi|_{y=\eta}
\end{equation}
and \begin{equation}\label{kin4}
\partial_n\phi|_{y=\eta}=0. \qquad
\end{equation}

We now replace $\phi$ by its harmonic conjugate $\psi$ such that
Cauchy-Riemann equations $\phi_x=\psi_y$ and $\phi_y=-\psi_x$ are
valid. BC  \e{kin4}  for Laplace Eq.
\begin{align} \label{psiBVP}
\nabla^2\psi=0 \
\end{align}
 at the free
surface reduces to vanishing of tangential  derivatives
$\partial_{\tau}\psi|_{y=\eta}=0$  because
$\partial_{\tau}\psi|_{y=\eta}=-\partial_n\phi|_{y=\eta}.$  Thus without the loss of
generality we set%
\begin{equation} \label{bcpsi}
\psi|_{y=\eta}=0.
\end{equation}
 BC at either $y\to-\infty$ or
$|x|\to \infty$ are reduced to%
\begin{equation} \label{bcpsiinfinity}
\psi\to-Vy\sqrt{\rho_n\rho_s}/\rho=-cy.
\end{equation}
If we introduce the stream
functions $\Psi_{n,s}$ for the components of He-II  (they satisfy
Cauchy-Riemann equations $\partial_x\Phi_{n,s}=\partial_y\Psi_{n,s}$
and $\partial_{y}\Phi_{n,s}=-\partial_x\Psi_{n,s}$), then
$\psi=(\Psi_n-\Psi_s)\sqrt{\rho _n\rho_s}/\rho$. Note that  $\psi$ is fully determined by $\eta(x,t)$  from \e{bcpsi} and \e{bcpsiinfinity} while being
independent  on $\Phi$. Dynamic BC \eqref{bern2} in terms of $\Phi$
and $\psi$ is given by
\begin{equation}\label{bern3}
\left .\frac{\partial\Phi}{\partial t}+\frac{(\nabla\Phi)^2}{2}
+\frac{(\nabla\psi)^2}{2}\right |_{y=\eta}=\frac{c^2}{2}-\frac{P_\alpha+P_g}{\rho}.
\end{equation}

Eqs.  \e{kin2}, \e{bcpsi} and \e{bern3} together with  $\nabla^2\Phi=\nabla^2\psi=0$
  and  BCs \e{bcPhiinfinity} and \e{bcpsiinfinity} at infinity form a closed set of equations equivalent (through harmonic conjugation) to solving two-fluid He-II hydrodynamics
  for KHI problem.
It is remarkable that  this set of equations  is equivalent (up to
trivial change of constants) to the problem of 2D dynamics of
charged surface of ideal fluid  in the limit when this surface
charges fully screen the electric field above the fluid free
surface. This limit was realized experimentally for He-II (with
negligible normal component) free surface charged by electrons
\cite{EdelmanUFN1980}. In that case  $\Phi$ has the meaning of the
only (ideal) fluid component and $\psi$ represents (up to the
multipication on constant) the electrostatic potential in the ideal
fluid. Also the term  $\propto(\nabla\psi)^2$ in Eq. \e{bern3}
corresponds to the electrostatic pressure.

Refs. \cite{Zubarev_JETPLett_2000, Zubarev_JETP_2002} found  exact time-dependent solutions for this problem of the dynamics of charged surface of superfluid He-II  in the limit of zero surface tension and gravity as well as  for the limit of zero temperature (i.e. neglecting the normal component of He-II). We now apply that approach for full (nonlinear) KHI problem with finite temperature by neglecting the last term in right-hand side %(r.h.s.)
of Eq. \eqref{bern3} as well as we provide  estimates of the applicability of such neglect of surface tension and gravity for two-component dynamics of He-II.

Our goal is to reduce Eqs.  \e{kin2}, \e{bcpsi} and \e{bern3} together with  $\nabla^2\Phi=\nabla^2\psi=0$
 and  BCs \e{bcPhiinfinity} and \e{bcpsiinfinity} to  solving LGE. Differentiation of Eq. \e{bcpsi} over  $t$ and $x$ results in
\[
\eta_t=-\left.\frac{\psi_t}{\psi_y}\right|_{y=\eta},
\qquad
\eta_x=-\left.\frac{\psi_x}{\psi_y}\right|_{y=\eta}
\]
respectively. Using these expressions in kinematic BC  \eqref{kin2} rewritten
in the equivalent form
\[
\eta_t=\left.\Phi_y-\eta_x\Phi_x\right|_{y=\eta},
\]
allows to obtain that\begin{equation}\label{kin5}
\left .\psi_t+\nabla\psi\cdot\nabla\Phi \right|_{y=\eta}=0.
\end{equation}
The sum and difference of Eqs. \eqref{bern3} and \eqref{kin5} (with $P_{\alpha}=P_g=0$) result in
\begin{equation}\label{para}
\left.F_t^{(\pm)} \mp cF_y^{(\pm)}
+(\nabla F^{(\pm)})^2\right|_{y=\eta}=0,
\end{equation}
where we introduced the harmonics potentials
\begin{equation}\label{pm_pot}
F^{(\pm)}=(\Phi\pm\psi\pm cy)/2,
\end{equation}
which satisfy the Laplace Eqs.  %
\begin{equation} \label{LaplFpm}
\nabla^2 F^{(\pm)}=0
\end{equation}
and decay at infinity,%
\begin{equation} \label{Fpmdecay}
F^{(\pm)}\to 0 \ \text{ for} \ y\to-\infty \ \text{or} \ |x|\to \infty.
\end{equation}
According to Eqs.  \e{bcpsi} and \e{pm_pot}, the motion of free surface is determined by the implicit expression
\begin{equation}\label{gran}
c\eta=\left .F^{(+)}-F^{(-)}\right |_{y=\eta}.
\end{equation}
Returning to physical velocity potentials $\Phi_{n,s}$ and stream functions $\Psi_{n,s}$, we find that
\begin{equation}\label{ff}
2\rho F^{(\pm )}=\rho_n\Phi_n+\rho_s\Phi_s
\pm\sqrt{\rho_n\rho_s}\left(\Psi_n-\Psi_s+Vy\right).
\end{equation}

Eqs. \e{LaplFpm}  together with BCs \e{para}, \e{Fpmdecay} and \e{gran}  are equivalent to the KHI problem.  It is crucial that nonlinear  Eqs. \eqref{para} decouple into separate Eqs. for $F^{(+)}$ and $F^{(-)}$.
We note that such decoupling does not occur for the classical KHI problem (the interface between two fluids) because in that case  the velocity potentials and stream functions of each of two fluids are defined in physically distinct regions ($y<\eta$ and $y>\eta$) thus making impossible a superposition of the type \eqref{ff}.
(Decoupling is however possible by other method in small angle %(small nonlinearity)
approximation which takes into account only leading quadratic nonlinearity in perturbation series for classical KHI \cite{Zubarev_Kuznetsov_JETP_2014}.)

The full set of equations   \e{para}, \e{LaplFpm}-\e{gran}  is still generally  coupled through Eq. \e{gran}. But an additional assumption (reduction) that either
\begin{equation} \label{Fpmddecouple}
 F^{(+)}=0 \ \text{or} \ F^{(-)}=0
\end{equation}
ensures the closed Eqs.   either for $F^{(+)}$ or $F^{(-)}$ which have a wide family of exact nontrivial solutions described below. Also that  assumption remains valid as time evolves. It follows from  Eqs. \e{pm_pot} that Eqs.  \e{Fpmddecouple} ensure  the  relations between $\Phi_{n,s}$ and $\Psi_{n,s}$ as
\[
\mp\sqrt{\rho_n\rho_s}\left(\Psi_n-\Psi_s+Vy\right)
=\rho_n\Phi_n+\rho_s\Phi_s,
\]
respectively.

We now comment on the physical meaning of  our reductions  \e{Fpmddecouple}, based on the particular limit  of small amplitude surface waves. In that limit we neglect the nonlinear term in Eq.  \e{para} which results in the linear system of equations which we solve in the form of plane waves
\begin{equation}\label{Flin}
\begin{split}
& F^{(\pm)}=a^{(\pm)}\exp(\I kx+ky-\I\omega^{(\pm)}t),
 \\
& \eta= b^{(+)}\exp(\I kx-\I\omega^{(+)}t)+ b^{(-)}\exp(\I kx-\I\omega^{(-)}t),
\end{split}
\end{equation}
where $a^{(\pm)}$ and $b^{(\pm)}$ are small constants,
$\omega^{(\pm)}$ are frequencies and $k$ is the wavenumber. First
Eq. in \e{Flin} ensures the exact solution of Eq.   \e{LaplFpm} with
decaying BC at  $y\to-\infty.$ Substitution of  \e{Flin} into Eq.
\e{gran} and the linearization of  \e{para} results in the
relations\begin{equation}\label{vetvi} \omega^{(\pm)}=\pm\I c k,
\qquad c b^{(\pm)}=\pm a^{(\pm)}.
\end{equation}
which are two branches of the dispersion relation of KHI with zero gravity and zero surface tension \cite{LandauLifshitzHydrodynamics1989,KorshunovEuriphysLett1991,KorshunovJETPLett2002}.  The branches with superscripts ``$+$''
and ``$-$'' correspond to exponentially growing and decaying perturbations of flat free surface, respectively.  Then Eqs.  \e{Fpmddecouple} choose one of these two branches. Thus Eqs. \e{para}, \e{LaplFpm}-\e{gran} together with \e{Fpmddecouple} represent the fully nonlinear stage of such separation into two branches.

The generic initial conditions include  both unstable and stable part \e{Flin} with the unstable part dominates as time evolves. Also it was shown in Refs.~\cite{Zubarev_JETP_2002,Zubarev_JETP_2008} that small  perturbation of  $F^{(-)}$ on the background of large $F^{(+)}$ decays to zero. Thus the choice of the reduction  $F^{(-)}=0$ (which is assumed below) in   Eq. \e{Fpmddecouple} is natural one to address the nonlinear stage of KHI.
Then Eqs.  \e{pm_pot} imply that  $F^{(+)}=\Phi=\psi+cy,$ i.e.    $\Phi$    is determined by $\psi$ and $\eta.$ The  boundary value problem (BVP) \e{psiBVP}-\e{bcpsiinfinity} solves for $\psi$ is at each $t.$ %
Then the  motion of free surface is determined by  Eq.  \eqref{kin2} as
\begin{equation}\label{kinred}
(\eta_t-c)(1+\eta_x^2)^{-1/2}
=\left.\partial_n\psi\right|_{y=\eta}.
\end{equation}
To solve BVP \e{psiBVP}-\e{bcpsiinfinity}  we consider the conformal map $z=z(w,t)$ \cite{DKSZ1996} from the lower complex half-plane  $-\infty<v\leq 0,$  $-\infty<u<+\infty$ of the complex variable $w=u+\I v$ into the area $-\infty<y\le\eta(x,t)$ occupied by the fluid in the physical plane $z=x+\I y$  with the real line $v=0$ mapped into fluid free surface.
Then the free surface is given in the parametric form $y=Y(u,t)\equiv\mbox{Im}\,z(u,t)$ and $x=X(u,t)\equiv\mbox{Re}\,z(u,t)$.
Solutions of both BVP  \e{psiBVP}-\e{bcpsiinfinity}  and the harmonically conjugated BVP $\nabla^2\phi=0,$ \e{phiBCinfinity}, \e{kin4} in $(u,v)$ variables are given by $\phi+\I\psi=-c(u+\I v$). It means that the conformal variables  $u$ and $v$ have a simple physical meaning:
$u=-\phi/c$ and $v=-\psi/c$ corresponding  (up to multipication to the constant $-1/c$) to the harmonically conjugated potentials $\phi$ and $\psi$.

We consider $w$ as independent variable while  $z(w,t)$ as the unknown function. Then Eq.    \eqref{kinred} is given by %$\begin{equation}\label{lge0}
$Y_tX_u-Y_u X_t=cX_u-c,
$ %\end{equation}
which can rewritten as
\begin{equation}\label{lge1}
\mbox{Im}\,\left(\bar{G}_t G_u\right)=c.
\end{equation}
where $G(u,t)=z(u,t)-\I ct$. Eq. \eqref{lge1} has the form of LGE, which is integrable in a sense of the existence of  infinite number of integrals of motion and relation to the dispersionless limit of the integrable
Toda hierarchy   \cite{MineevWeinsteinWiegmannZabrodinPRL2000}.

LGE has the infinite number of exact solutions often involving
logarithms (see e.g.
Refs.\cite{ShraimanBensimonPRA1984,HowisonSIAMJApplMath1985,BensimonKadanoffLiangShraimanTangRevModPhys1986,MineevDawsonPRE1994}).
We consider a  periodic solution
\cite{BensimonKadanoffLiangShraimanTangRevModPhys1986} with the
wavenumber $k$ as\begin{equation}\label{resh}
z=w-\I kA^2(t)/2-\I A(t)e^{-\I kw},
\end{equation}
where $A(t)$ is the amplitude of free surface surface perturbation satisfying a nonlinear  ordinary differential equation
%\[
$dA/dt=ckA\left(1-k^2A^2\right)^{-1}
$%\]
  which develops a finite-time singularity in $dA/dt$ at the time $t=t_c$ when  $A(t_c)=1/k$.
As $t$ approaches $t_c$,  a leading order solution is given by $A= 1/k-\sqrt{c\tau/k}+O(\tau)$, where $\tau=t_c-t$.
Singularities of the conformal map \e{resh} are determined by a condition $z_w=0$ implying that they approach the real line  $v=0$ from above with the increase of $t$ until reaching $v=0$ at $\tau=0$ and $u=2\pi n/k$, $n=0,\pm1,\pm2,\ldots$  In particular, choosing $n=0,$  expanding at  $u=0$ and assuming $\tau \to 0$ we obtain that
\begin{equation}\label{xyser}
\begin{split}
& X(u,\tau)= u\sqrt{ck\tau}+\frac{k^2u^3}{6}+O(u\tau^{})+O(u^3\tau^{1/2}),
 \\
& Y(u,\tau)= \frac{-3}{2k}+2\sqrt{\frac{c\tau}{k}}+\frac{ku^2}{2}+O(\tau^{})+O(u^2\tau^{1/2}).
\end{split}
\end{equation}
Fig.~1 shows an example of such solution at different  $t$.
It follows from Eq.~\eqref{xyser} that
a cusp pointing downward (a dimple) $y+3/2k\propto|x|^{2/3}$ is formed at the free surface  at $t=t_c$ (i.e. $\tau=0$) with the vertical velocity diverging as $\tau^{-1/2}$ at the tip of the cusp \cite{ShraimanBensimonPRA1984,BensimonKadanoffLiangShraimanTangRevModPhys1986}.

\begin{figure}[h]
\center{\includegraphics[width=0.8\linewidth]{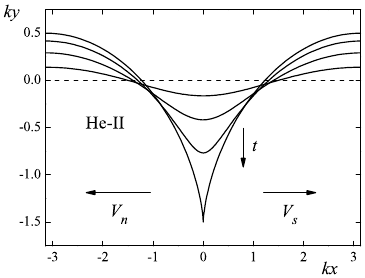}}
\caption{Evolution of an initial periodic perturbation of the free
surface $y(x)$  for Eq. \e{resh} with $kA(0)\approx 0.15$. The
surface shape is shown over one spatial period for the times
$ckt=0,\,0.8,\,1.2,\,1.4$ until the cusp singularity is formed. The
dashed line shows the unperturbed free surface, $y\equiv0$.}
\end{figure}

Near the  singularity (the tip of the cusp) one has to take into account the surface tension and the finite viscosity of the normal component to regularize the singularity.
Surface tension near the singularity is given
by $P_{\alpha}\approx\alpha/r$, where $r$ is the radius of
curvature of free surface. It follows from Eq. \eqref{xyser}  that
$r\approx c\tau$, which implies that
$P_{\alpha}\approx\alpha/c\tau$. At the same time, the dynamic
pressure $P_v$ which determines the development of  KHI in LGE
reduction is given by $P_v=\rho
v^2/2\equiv\rho\left[(\nabla\Phi)^2)+(\nabla\psi)^2\right]/2$, where
$v$ is the typical velocity. Near singularity
$v\simeq\sqrt{2c/k\tau}$ and $P_v=\rho c/k\tau$. Thus both $P_v$ and
$P_{\alpha}\propto \tau^{-1}$. Surface tension effect is small if the Weber number $\mbox{We}=P_v/P_{\alpha}$, the ratio of dynamic
and surface tension pressures, is well above 1. Using $\mbox{We}\approx\rho c^2/\alpha k=\rho_n\rho_sV^2/(\rho\alpha
k)$, and assuming We$\gtrsim1$ for applicability of  LGE regime, we obtain the condition for  the wavelength
$\lambda=2\pi/k\gtrsim2\pi\rho\alpha/(\rho_n\rho_s V^2)$.
 He-II at the temperature 1.5~K has $\rho_n=0.016\;\text{g}/\text{cm}^3$, $\rho_s=0.129\;\text{g}/\text{cm}^3$ and $\alpha=0.332\;\text{dyn}/\text{cm}$ \cite{DonnellyBarenghiJPhysChemRefData1998}. E.g. if $V=50\;\text{cm}/\text{s}$ then $\lambda\gtrsim0.58$ mm which provides a good range of values  below a gravity-capillary length  $2\pi\sqrt{\alpha/\rho g}\approx 3.03$~mm.
Increasing $V$  further extends the range of applicability of LGE regime.

The relative strength of inertial and viscous forces near the singularity is determined by the Reynolds number given by  $\mbox{Re}=vr/\nu$, where $\nu$ is kinematic viscosity He-II. Using that $v\approx\sqrt{2c/k\tau}$ and $r\approx c\tau$,
implies that $\mbox{Re}\approx c\nu^{-1}\sqrt{2r/k}$, i.e. Re turns small for $r\to 0$ and viscosity has to be taken into account. A typical scale $r_\nu$ below which the flow of the normal component cannot be considered as potential  one is estimated by setting $\mbox{Re}\approx 1$ which gives $r_\nu\approx k\nu^2/2c^2$. For the temperature 1.5~K, we use $\nu=9.27\cdot 10^{-5}\;\text{cm}^2/\text{s}$ \cite{DonnellyBarenghiJPhysChemRefData1998}.
Then $r_\nu\approx 1.8\cdot 10^{-10}$~cm, i.e. $r_\nu\ll \lambda$ thus the viscous effect is much less than the surface tension.
The influence of gravity, which is determined by the Froude number $\mbox{Fr}=P_v/P_g$, is expected to be also small near the
singularity. During singularity formation the gravity pressure $P_g\simeq \rho gy$ is finite while $P_v$ diverges as $\tau^{-1}$ implying divergence of Froude number, so the gravity effects can be neglected.

We conclude that we reduced the nonlinear KHI dynamics to the solution of LGE which has the infinite set of exact solutions with the generic formation of cusps at the free surface in a finite time. That LGE regime is applicable down the the capillary spatial scale. The capillarity   has to be taken into account only to regularize the singularity at the small spatial scale.

%\begin{acknowledgments}
{\it Acknowledgments.} The authors thank L.P. Mezhov-Deglin for helpful discussions. The  work of P.L.  was partially supported by the National Science Foundation  DMS-1412140.
The work of P.L. on KHI was supported by the Russian Scientific Foundation, Grant No. 14-22-00259. The  work of  N.Z. was supported by the Russian Foundation for Basic Research (grant nos. 16-08-00228 and 17-08-00430) and by the Presidium of the Ural Branch of the RAS (grant no. 15-8-2-8).
%\end{acknowledgments}

%%%%%%%%%%%%%%%%%%%%%%%%%%%%%%%%%%%%%%%%%%%%%%%%%%

%\bibliographystyle{jpp}
% \bibliography{biblionls,surfacewaves,lushnikov,helium}

\end{document}